\documentclass[letter,twocolumn]{jpsj3}
\usepackage{txfonts}
\usepackage{braket}
\usepackage{color}
\usepackage{enumerate}
\usepackage{fancyhdr}
\usepackage{color}

\title{Electric potentials and field lines for uniformly-charged tube and cylinder expressed by Appell's hypergeometric function and integration of $Z(u|m) \mathrm{sc}(u|m)$}

\author{Daisuke A. Takahashi\thanks{daisuke.takahashi@keio.jp}}
\inst{Research and Education Center for Natural Sciences, Keio University, \\ Hiyoshi 4-1-1, Yokohama, Kanagawa 223-8521, Japan} %\\
\abst{The closed-form expressions of electric potentials and field lines for a uniformly-charged tube and cylinder are presented using elliptic integrals and Appell's hypergeometric functions, where field lines are depicted by introducing the concept of the field line potential in axisymmetric systems, whose contour lines represent electric field lines outside the charged region, thought of as an analog of the conjugate harmonic function in the presence of non-uniform metric. The field line potential for the tube shows a multi-valued behavior and enables us to define a topological charge. The integral of $Z(u|m)\operatorname{sc}(u|m)$, where $ Z $ and $ \operatorname{sc} $ are the Jacobi zeta and elliptic functions, is also expressed by Appell's hypergeometric function as a by-product, which was missing in classical tables of formulas. \\ \indent In the Addendum appended after the main article, several relevant references are provided and the decomposition of the solution by ``degrees of transcendence'' is proposed. \\ \indent The attached supplementary calculations provide detailed derivations of several formulas including the integral in the title and discuss the resemblance between the field line potential for tube and the electric potential for disk.}
\begin{document}

%%% references output based on chapterbib %%%%%%%%%%%%%%%%%%%%%%%%%
%\include{EPandEFLforCylandTube-ver4-main}
%\fancyfoot[C]{\footnotesize \thepage}
%\include{EPandEFLforCylandTube-ver4-addenda}
%\include{EPandEFLforCylandTube-ver5-suppl}
%%% run the following %%%%%%%%%%%%%%%%%%%%%%%%%%%%%%%%%%%%%%%%%%%%%%%%
%% First, run the following with uncomment \usepackage{chapterbib}
%% platex EPandEFLforCylandTube-ver5
%% bibtex EPandEFLforCylandTube-ver4-main
%% bibtex EPandEFLforCylandTube-ver4-addenda
%% bibtex EPandEFLforCylandTube-ver5-suppl
%% platex EPandEFLforCylandTube-ver5
%% platex EPandEFLforCylandTube-ver5
%% Next, paste the content of .bbl files to the .tex files.
%% Third, delete duplicated references of ver5-suppl.bbl., comment chapterbib (to cite the references of ver4-main.bbl in ver5-suppl.tex ), and run platex.
%%
%%% c.f.: https://tex.stackexchange.com/questions/333617/how-to-use-chapterbib-package-syntax %%%
%
%%%%%%%%%%%%%%%%%%%%%%%%%%%%%%%%%%%%%%%%%%%%%%%%%%%%%%%%%%%%%%%%%%%%%%%%%%%%%%%%%%%%%%%
%%% (for arxiv submission) the paste from EPandEFLforCylandTube-ver4-main.tex, EPandEFLforCylandTube-ver4-main.bbl, EPandEFLforCylandTube-ver4-addenda.tex, EPandEFLforCylandTube-ver4-addenda.bbl, EPandEFLforCylandTube-ver5-suppl.tex, EPandEFLforCylandTube-ver5-suppl.bbl, with some manual modification of bibitem's mentioned above. %%%%%%%%%%%%%%%%

\maketitle

%\section{Introduction}

\indent \textit{Introduction ---} Multi-variable generalizations of hypergeometric functions such as Appell's hypergeometric functions \cite{ErdelyiHTF1,KimuraHFTV1973}, the Kamp\'e de F\'eriet functions, and the Lauricella-Saran functions \cite{Exton,SrivastavaKarlsson}, emerge in modern physics in a variety of ways; the example includes the Feynman integrals \cite{Feng2018,Bera2025}, the capacity of entanglement \cite{Bhattacharjee2021}, and the loop amplitude in photon-photon scattering \cite{Bera2025}. The Mathematica packages for these functions are constructed as well \cite{Ananthanarayan2023,Bera2025}. Those on a finite field have also been studied in number theory \cite{Ito2023}. We should also mention the recent developments on the elliptic generalization in mathematical physics \cite{Spiridonov2008}. It is not of academic interest that the solutions of physical problems can be written by these special functions, because their linear transformation formulae, integral representations, and differential equations enable us to predict their global behaviors beyond the definition series which only has a finite radius of convergence. \\ 
\indent Here, we report an application of Appell's hypergeometric function to rather an elementary problem. That is, we provide closed-form expressions for electric potentials and field lines for a uniformly-charged tube and cylinder.
We mention that the potentials for ring and disk written by elliptic integrals are well known in astronomy and widely used \cite{Lass1983,10.1046/j.1365-8711.2000.03523.x,Fukushima2010,Fukushima2016}. \\
\indent In the integration of the above problem, after separating the terms expressible by elementary and elliptic integrals, only the following term remains: 
\begin{align}
	I_{\text{hyg}}(m,A;\theta)\coloneqq \int_0^\theta\! d\theta\tanh^{-1}\frac{A}{\sqrt{\smash[b]{1-m\sin^2\frac{\theta}{2}}}}. \label{eq:Ihygdef}
\end{align}
	This integral does not reduce to an abelian integral by any change of variable, so it cannot be expressed by the Riemann theta functions which are used to express finite-zone solutions in classical integrable systems \cite{TanakaDate,BelokolosBobenkoEnolskiiItsMatveev}. However, since the parameter derivatives $ \frac{\partial I_{\text{hyg}}}{\partial A} $ and $ \frac{\partial I_{\text{hyg}} }{\partial m} $ are written by elliptic integrals, $ I_{\text{hyg}} $ allows a double integral expression of the algebraic function, implying that it could possibly be expressed by some multi-variable hypergeometric functions. Indeed, using Appell's hypergeometric function of the second kind \cite{ErdelyiHTF1}
	\begin{align}
		F_2^{\text{Appell}} \left( \begin{smallmatrix}\alpha;\beta,\beta' \\ \gamma,\gamma'\end{smallmatrix};x,y \right)\coloneqq \sum_{j,l=0}^\infty \frac{(\alpha)_{j+l}(\beta)_j(\beta')_l}{j!l!(\gamma)_j(\gamma')_l}x^jy^l, \label{eq:AppellF2}
	\end{align}
	where $ (x)_n=x(x+1)\dots(x+n-1) $ is the Pochhammer symbol, the above integral, writing $ s=\sin\frac{\theta}{2} $, is given by
	\begin{align}
		&I_{\text{hyg}}(m,A;\theta)=\pi A \operatorname{sgn}(s) F_2^{\text{Appell}}\left( \begin{smallmatrix}\frac{1}{2};\frac{1}{2},1 \\ 1,\frac{3}{2}  \end{smallmatrix};m,A^2 \right)\notag \\
		&\quad-2As\sqrt{1-s^2}\sum_{k=0}^\infty \frac{(1)_k}{(\frac{3}{2})_k}(1-s^2)^k F_2^{\text{Appell}}\left( \begin{smallmatrix} \frac{1}{2};1+k,1 \\ 1,\frac{3}{2}  \end{smallmatrix};ms^2,A^2 \right). \label{eq:IhygF2}
	\end{align}
	In particular, the definite integral $ I_{\text{hyg}}(m,A;\pi) $ is expressed only by the first term and hence the above-mentioned problem can be solved. \\
	\indent Furthermore, rewriting the elliptic integral of the third kind using the Jacobi zeta and theta functions \cite{Todaelliptic,Akhiezer}, we provide an integral of $Z(u|m)\operatorname{sc}(u|m)$, which was missing in classical table of formulas for elliptic functions \cite{ByrdFriedman,AbramowitzStegun}. Thus, even a fundamental problem in classical electromagnetism could sometimes offer an opportunity to improve our knowledge on higher transcendental functions. \\
	\indent The paper is organized as follows. First, we introduce the concept of the field line potential in axisymmetric systems whose contour line describes the electric field line. We in particular emphasize that its definition is restricted to the chargeless regions. Next, we provide and summarize the exact expressions of electric potentials and field line potentials for a uniformly-charged cylinder and tube. We also point out that, for the case of tube, the field line potential is multi-valued and has a topological charge. Lastly, we prove Eq.~(\ref{eq:IhygF2}), and also provide a by-product integration formula for a product of the Jacobi zeta and elliptic functions, which was missing in classical tables of formulas for elliptic functions \cite{ByrdFriedman,AbramowitzStegun}. \\
	\indent Throughout the paper, we follow the notations of elliptic integrals and functions in Ref.~\citen{AbramowitzStegun} and Mathematica. So, the incomplete elliptic integrals of the first, the second, and the third kind are denoted by $F(\varphi|m),\ E(\varphi|m)$, and $\Pi(n;\varphi|m)$.

\begin{figure*}[tb]
\begin{tabular}{lll}
(a)&(b)&(c) \\
\includegraphics[width=0.32\linewidth]{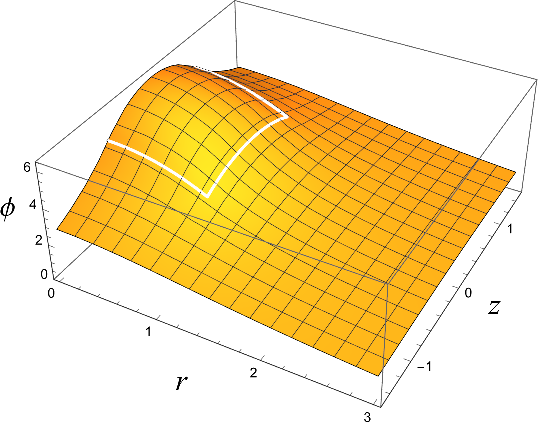}&\includegraphics[width=0.32\linewidth]{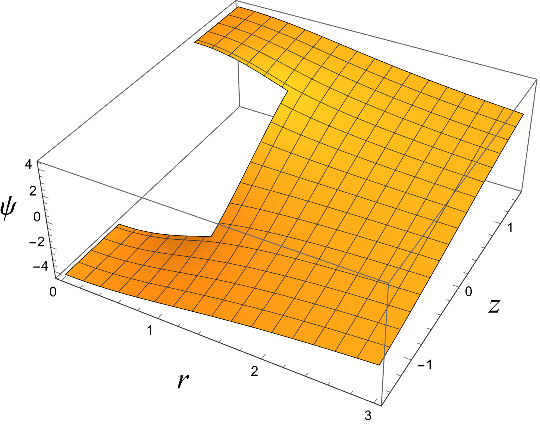}&\includegraphics[width=0.25\linewidth]{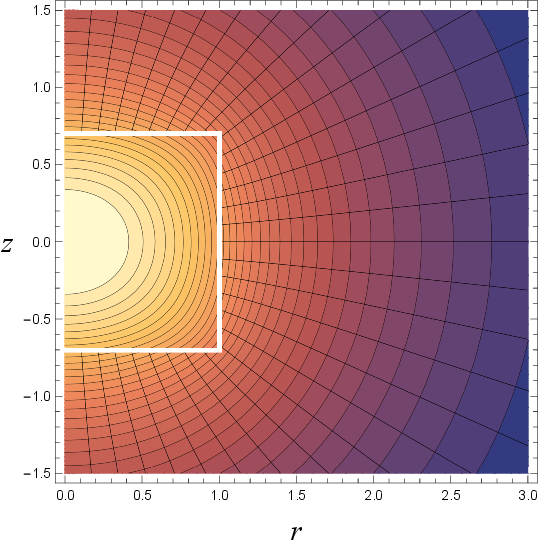} %\\
\end{tabular}
\caption{The electric potential and field line potential made by a uniformly charged cylinder. Figures (a) and (b) represent $ \phi^{\text{cyl}} $ and  $ \psi^{\text{cyl}} $ with parameters $ \rho_0=1,\ R=1, $ and $ Z=0.7 $. The potential $ \phi^{\text{cyl}} $ is $C^1$ on the surface. Figure (c) is a contour plot, showing that equipotential lines of $ \phi $ and $ \psi $ are orthogonal. Inside the cylinder, $ \psi $ is not given by Eq.~(\ref{eq:psidef}), so we leave it unplotted, though the field lines themselves do exist.}
\label{fig:cyl}
\end{figure*}

\begin{figure*}[tb]
\begin{tabular}{lll}
(a)&(b)&(c) \\
\includegraphics[width=0.32\linewidth]{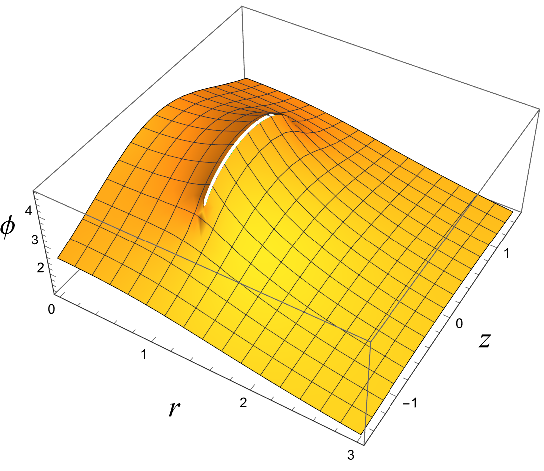}&\includegraphics[width=0.32\linewidth]{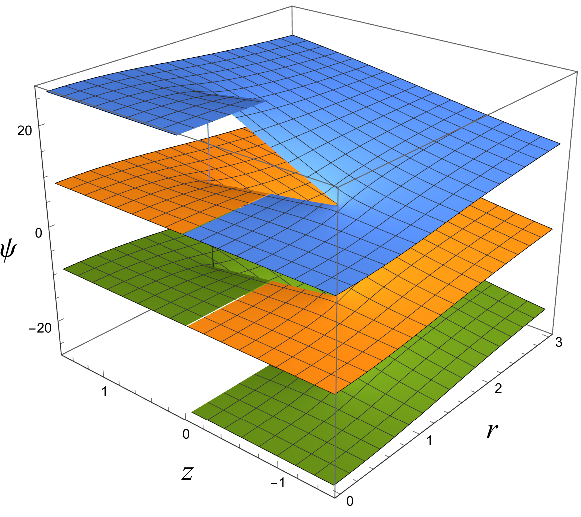}&\includegraphics[width=0.25\linewidth]{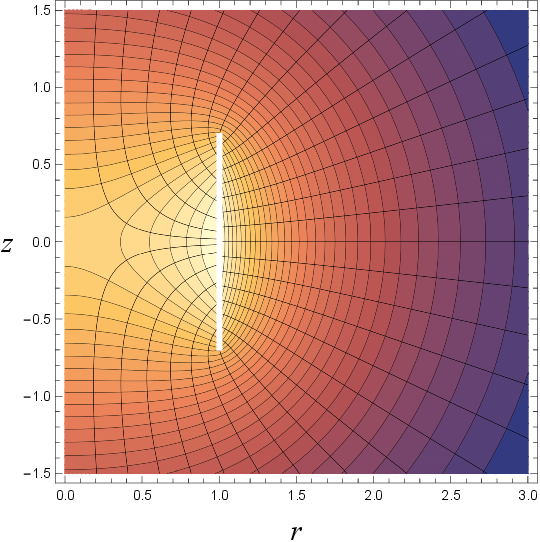}
\end{tabular}
\caption{The electric potential and field line potential made by a uniformly charged tube. We set parameters $ \sigma_0=1,\ R=1, $ and $ Z=0.7 $.  Figure (a) represent $ \phi^{\text{tube}} $, which has a non-differentiable corner at $ r=R,\ |z|<Z $. In figure (b), we plot $ \psi^{\text{tube}}+n\Delta\psi^{\text{tube}} $ with $ n=0,\pm1 $, showing the log-like multi-valued character.  Figure (c) is a contour plot.}
\label{fig:tube}
\end{figure*}

	\indent \textit{Field line potential ---} Before going to the main subject of this paper, we use a few paragraphs to introduce the field line potential whose equipotential lines coincide with the electric field lines. 
	We write the cylindrical coordinate as $ (r,\theta,z) $ and consider axisymmetric charge distribution $ \rho(r,\theta,z)=\rho(r,z) $. With the unit choice $ 4\pi\epsilon_0=1 $, the electric potential at position $ \boldsymbol{r}=(r\cos\theta,r\sin\theta,z) $ is given by
	\begin{align}
		\phi(r,z)=\iiint\frac{\rho(r',z')r'dr'd\theta' dz'}{L},\quad L=|\boldsymbol{r}'-\boldsymbol{r}|.
	\end{align}
	Henceforth, without losing generality, the angle of the observation point is chosen $ \theta=\pi $ by axisymmetry. Then, the field line potential $ \psi $ whose contour lines represent the electric field lines \textit{defined only outside the charged region} is given by
	\begin{align}
		\psi(r,z) &=\iiint \frac{r(z-z')(r+r'\cos\theta')}{L[L^2-(z'-z)^2]} \rho(r',z')r'dr'd\theta' dz' \notag \\
		&\quad + \text{(correction originating from a uniform field)},   \label{eq:psidef}
	\end{align}
	\textit{if $\rho(r,z)=0$.} Below we derive Eq.~(\ref{eq:psidef}), and explain why this $ \psi $ is valid only for the chargeless region.\\
	\indent While the electric field line is originally defined by a streamline of the electric field, i.e., the solution of the ordinary differential equation (ODE) $ \frac{\mathrm{d}}{\mathrm{d}s}\boldsymbol{x}(s)=\boldsymbol{E}(\boldsymbol{x}(s)) $, we follow another formulation. Let us consider a coordinate transformation $ u=\phi(r,z) $ and $ v=\psi(r,z) $ with $ \theta $ fixed. Since the equipotential line of the electric potential $ \phi $ and the field lines are everywhere orthogonal, the desired function $ \psi $ must be chosen so that the off-diagonal element of the metric tensor in the new coordinate vanishes: $ g_{uv}=0 \ \leftrightarrow \ \phi_r\psi_r+\phi_z\psi_z=0 $. Let us assume the form $ \psi_r= f(r)\phi_z,\ \psi_z=-f(r)\phi_r $. Using the Laplace equation outside the charged region $ \phi_{rr}+\frac{1}{r}\phi_r+\phi_{zz}=0 $ and the compatibility $ \phi_{rz}=\phi_{zr} $, we conclude that $f(r)=r$, that is, 
	\begin{align}
		\psi_r=r\phi_z,\quad \psi_z=-r\phi_r, \label{eq:axialCR}
	\end{align}
	and $ \psi $ satisfies the partial differential equation (PDE) given later [Eq.~(\ref{eq:Faraday})]. Integrating the latter of Eq. (\ref{eq:axialCR}), we find $ \psi=-\int dz r \frac{\partial \phi}{\partial r}$, which yields Eq.~(\ref{eq:psidef}). %, where without losing generality we set $ \theta=\pi $ by axisymmetry. 
	The second term of Eq. (\ref{eq:psidef}) complements the lost information on a uniform electric field, since the formula is obtained by an integration of the derivative $ \phi_r $.
	Note that the equivalence of the ODE-based streamline formulation and this PDE-based method can be traced back to the method of characteristic curves \cite{CourantHilbertII}. \\
	\indent Let us write the electric field $ \boldsymbol{E}=-\nabla\phi $ and its conjugate field $ \boldsymbol{\mathcal{B}}=\nabla\psi $. The field $ \boldsymbol{\mathcal{B}} $ has a mathematical property similar to magnetic fields, but not a real magnetic field. Taking into account the componentwise relation of Eq. (\ref{eq:axialCR}), $ E_r=\frac{1}{r}\mathcal{B}_z,\ -E_z=\frac{1}{r}\mathcal{B}_r $, their fundamental laws are given by
	\begin{alignat}{2}
		\nabla\cdot\boldsymbol{E}&=\frac{(\nabla\times \boldsymbol{\mathcal{B}})_\theta}{r}&&=-(\phi_{rr}+\tfrac{1}{r}\phi_r+\phi_{zz})=4\pi\rho, \label{eq:Gauss} \\
		(\nabla \times \boldsymbol{E})_\theta&=\ \nabla \cdot\left( \frac{\boldsymbol{\mathcal{B}}}{r} \right)&&=\psi_{rr}-\tfrac{1}{r}\psi_r+\psi_{zz}=0. \label{eq:Faraday}
	\end{alignat}
	Equation (\ref{eq:Gauss}) explains why we cannot use Eq.~(\ref{eq:psidef}) inside a charged region $ \rho\ne0 $ --- the potential function is introduced only for a rotation-free vector field. 
	
	\indent Here we give further remarks (i)-(v) about $ \psi $. (i) $ \psi $ makes sense only for axisymmetric systems. For non-symmetric $ \rho $, the path of each field line is not closed in one plane, and hence no natural choice for remaining two coordinates perpendicular to $ \phi $ exists. (ii) $ \psi $ is possibly a multivalued function, though its gradient $ \nabla \psi $ is always a single-valued vector field. (iii) The concept of the electric field line itself does survive even inside the charged region $ \rho\ne 0 $; what we only claim here is that the function satisfying the relation $ \phi_r\psi_r+\phi_z\psi_z=0 $ cannot be given by Eq.~(\ref{eq:psidef}). (iv) $ \psi $ satisfies the same superposition principle as $ \phi $; that is, if $ \psi_i,\ i=1,2, $ are the solution for the axisymmetric charge profile $ \rho_i $, then $ \psi_1+\psi_2 $ is the solution for $ \rho_1+\rho_2 $. For example,  $ \psi $ for a point charge $ q $ at origin is easily found to be $ \psi=\frac{qz}{\sqrt{r^2+z^2}} $, and hence $ \psi $ for two point charges $ q_\pm $ located at $ (r,z)=(0,\pm a) $ is given by their superposition: $ \psi=\sum_\pm \frac{q_\pm (z\mp a)}{\sqrt{r^2+(z\mp a)^2}} $. Drawing the equipotential line of this $ \psi $, we soon find a familiar picture of the electric field lines created by two point charges found in textbooks. (v) In the two-dimensional electromagnetism, the field line potential $ \psi $ is just a conjugate harmonic function of $ \phi $ and they satisfy the Cauchy-Riemann relation, a counterpart of Eq.~(\ref{eq:axialCR}); the example of single point charge is $ \phi=-\log r=-\ln\sqrt{x^2+y^2} $ and $ \psi=\theta=\tan^{-1}\frac{y}{x} $ and the latter is indeed multi-valued. The present formulation for axisymmetric three-dimensional systems can therefore be regarded as a generalization with non-uniform metric.

	\indent \textit{Cylinder ---} Now let us go to the uniformly-charged cylinder. Writing $ L=\sqrt{r^2+r_0^2+2rr_0\cos\theta+z^2} $ and $ L_0=[L]_{\theta=0}=\sqrt{(r+r_0)^2+z^2} $, the triple indefinite integrals
	\begin{align}
		I^{\text{cyl}}(r,\theta,z;r_0)&=\iiint\frac{rdrd\theta dz}{L},\\ 
		J^{\text{cyl}}(r,\theta,z;r_0)&=\iiint\frac{-r_0z(r_0+r\cos\theta)rdrd\theta dz}{L(L^2-z^2)}
	\end{align}
	are given by $ I^{\text{cyl}}(r,\theta,z;r_0)=\sum_{\text{T}=\text{trig,ell,hyg}}I^{\text{cyl}}_{\text{T}}(r,\theta,z;r_0) $ and $ J^{\text{cyl}}(r,\theta,z;r_0)=\sum_{\text{T}=\text{trig,ell}}J^{\text{cyl}}_{\text{T}}(r,\theta,z;r_0) $, where 
	\begin{align}
		&I_{\text{hyg}}^{\text{cyl}}= \tfrac{r^2}{2}I_{\text{hyg}}(m,A;\theta), \\
		&I_{\text{trig}}^{\text{cyl}}=\tfrac{-r_0^2\sin2\theta}{4}\tanh^{-1}\!\tfrac{z}{L}-zr_0\sin\theta\tanh^{-1}\tfrac{r+r_0\cos\theta}{L} \notag \\
		&\qquad\qquad+\tfrac{r_0^2\cos2\theta}{4} \tan^{-1}\tfrac{Lr_0\sin\theta}{z(r+r_0\cos\theta)},\\
		&I_{\text{ell}}^{\text{cyl}}=\tfrac{-3z(r_0^2+z^2)}{4L_0}F(\tfrac{\theta}{2}|m)+\tfrac{3zL_0}{4}E(\tfrac{\theta}{2}|m)+\tfrac{zr^2(r-r_0)}{4L_0(r+r_0)}\Pi\Big(\tfrac{4rr_0}{(r+r_0)^2};\tfrac{\theta}{2}\Big|m\Big)\notag \\
		&\qquad+\tfrac{z(2z^2-r_0^2)}{4L_0}\sum_{\alpha=\pm}\left[ 1\!-\!\tfrac{n_\alpha}{2}\left( 1\!+\!\tfrac{r}{r_0} \right) \right]\Pi(n_\alpha;\tfrac{\theta}{2}|m), \label{eq:Icylell} \\
		&J_{\text{trig}}^{\text{cyl}}=\tfrac{(3r_0^3-4r_0z^2)\sin\theta-r_0^3\sin3\theta}{8}\tanh^{-1}\tfrac{r+r_0\cos\theta}{L}-\tfrac{r_0^2z\sin2\theta}{2}\tanh^{-1}\!\tfrac{z}{L} \notag \\
		&\qquad +\tfrac{r_0^2 z\cos2\theta}{2}\tan^{-1}\tfrac{Lr_0\sin\theta}{z(r+r_0\cos\theta)}+\tfrac{Lr_0\sin\theta(-r+3r_0\cos\theta)}{6}, \\
		&J_{\text{ell}}^{\text{cyl}}=\tfrac{L_0[z^2-2(r^2+r_0^2)]}{6}E(\tfrac{\theta}{2}|m)+\tfrac{2(r^2-r_0^2)^2+z^2(r_0^2-2r^2-z^2)}{6L_0}F(\tfrac{\theta}{2}|m)\notag \\
		&\qquad+\tfrac{z^2r^2(r-r_0)}{2L_0(r+r_0)}\Pi \Big(\tfrac{4rr_0}{(r+r_0)^2};\tfrac{\theta}{2}\Big|m\Big)\notag \\
		&\qquad\quad-\tfrac{r_0^2z^2}{2L_0}\sum_{\alpha=\pm}\left[ 1\!-\!\tfrac{n_\alpha}{2}\left( 1\!+\!\tfrac{r}{r_0} \right) \right]\Pi(n_\alpha;\tfrac{\theta}{2}|m), \label{eq:Jcylell}
	\end{align}
	and $ m=\frac{4rr_0}{L_0^2},\ A=\frac{z}{L_0},\ n_\pm=\frac{2r_0}{r_0\pm\sqrt{\smash[b]{r_0^2+z^2}}}, $ and $ I_{\text{hyg}} $ is given by (\ref{eq:Ihygdef}) or (\ref{eq:IhygF2}). We bequeath another expression for the last terms in Eqs. (\ref{eq:Icylell}) and (\ref{eq:Jcylell}): $ 1\!-\!\tfrac{n_\alpha}{2}\Big( 1\!+\!\tfrac{r}{r_0}\Big) = \frac{L_0}{2r_0}s_\alpha\sqrt{n_\alpha(n_\alpha-m)} $ with $ s_\pm=\operatorname{sgn}\big(\sqrt{\smash[b]{r_0^2+z^2}}\mp r\big) $, which plays a key role in determination of $\phi_{\text{corr}}^{\text{cyl}}$ and $ \psi_{\text{corr}}^{\text{cyl}}$ (see below) using the formula 117.02 in Ref.~\citen{ByrdFriedman}. The other useful relation for definite integral $(\theta=\pi) $ is $ \sum_{\alpha=\pm}\left[ 1\!-\!\tfrac{n_\alpha}{2}\left( 1\!+\!\tfrac{r}{r_0} \right) \right]\Pi(n_\alpha|m)=K(m)+\tfrac{r-r_0}{r+r_0}\Pi \Big(\tfrac{4rr_0}{(r+r_0)^2}\Big|m\Big)+\frac{\pi L_0}{|z|}H(r_0-r), $ 
%	\begin{align}
%		\sum_{\alpha=\pm}\left(1-\frac{n_\alpha}{2}(1+\frac{r}{r_0})  \right)\Pi(n_\alpha|m)=K(m)+\frac{r-r_0}{r+r_0}\Pi\Big(\frac{4rr_0}{(r+r_0)^2}\Big|m\Big)+\frac{\pi L_0}{|z|}H(r_0-r),
%	\end{align}
	which is proved by equating the two expressions for the potential of the uniformly-charged disk calculated via $ \partial I^{\text{cyl}}/\partial z $ and Ref.~\citen{Lass1983}.\cite{Foot02} \\
	\indent The electric potential $ \phi^{\text{cyl}}(r,z) $, made by a cylinder of radius $R$ and height $2Z$ with uniform charge density $ \rho_0 $, is then constructed as follows. If the above indefinite integral formula had no artificial singularity, it would be $ \phi=\rho_0\Big[\big[2I^{\text{cyl}}(r',\pi,z'-z;r)\big]_{r'=0}^{r'=R}\Big]_{z'=-Z}^{z'=Z} $. However, we actually must include the contributions from the branch shift of the multivalued functions $ \tan^{-1} $ and $ \Pi $. The same also holds for the field line potential $ \psi^{\text{cyl}}(r,z) $, but it has one more correction shown in Eq.~(\ref{eq:psidef}). After careful determination of these corrections, we get $ \phi^{\text{cyl}}=\phi^{\text{cyl}}_{\text{hyg}}+\phi^{\text{cyl}}_{\text{ell}}+\phi^{\text{cyl}}_{\text{\vphantom{hl}corr}} $ and $ \psi^{\text{cyl}}=\psi^{\text{cyl}}_{\text{ell}}+\psi^{\text{cyl}}_{\text{\vphantom{hl}corr}} $, where
	\begin{align}
		\phi^{\text{cyl}}_{\text{T}}&=\rho_0\sum_{\beta=\pm 1} 2\beta I^{\text{cyl}}_{\text{T}}(R,\pi,\beta Z-z;r),\quad \text{T}=\text{hyg,\ ell}, \label{eq:cyl01} \\
		\phi^{\text{cyl}}_{\text{corr}}&=\pi\rho_0\Big[ r^2H(r-R)-2(z^2+Z^2) \Big]H(Z-|z|) \notag \\
		&\qquad\qquad -4\pi\rho_0|z|H(|z|-Z), \label{eq:phicylcorr} \\
		\psi^{\text{cyl}}_{\text{ell}}&=\rho_0\sum_{\beta=\pm 1}2\beta J^{\text{cyl}}_{\text{ell}}(R,\pi,\beta Z-z;r), \\
		\psi^{\text{cyl}}_{\text{corr}}&=-2\pi \rho_0r^2 z H(Z-|z|) \notag \\
		&\quad +2\pi \rho_0Z \operatorname{sgn}(z) \Big[ -r^2+R^2H(R\!-\!r)\Big]H(|z|-Z), \label{eq:cyl04}
	\end{align}
	and $H(x)$ is the Heaviside function. We can check $ \phi^{\text{cyl}} \to \frac{Q}{\sqrt{r^2+z^2}} $ and $ \psi^{\text{cyl}} \to \frac{Qz}{\sqrt{r^2+z^2}} $ with total charge $ Q=2\pi R^2Z\rho_0 $ at spatial infinity $ \sqrt{r^2+z^2}\to \infty $. The plots are shown in Fig.~\ref{fig:cyl}.

	\indent \textit{Tube ---} Next, we consider the tube whose thickness is negligible. The double indefinite integral formulae are given by
	\begin{align}
		&I^{\text{tube}}(r,\theta,z;r_0)=\iint\frac{d\theta dz}{L}=I_{\text{hyg}}(m,A;\theta) ,\\ 
		&J^{\text{tube}}(r,\theta,z;r_0)=\iint\frac{-r_0z(r_0+r\cos\theta)d\theta dz}{L(L^2-z^2)} \notag \\
		&\qquad=\tfrac{r^2-r_0^2}{L_0}F(\tfrac{\theta}{2}|m)-L_0E(\tfrac{\theta}{2}|m)+\tfrac{z^2}{L_0}\tfrac{r-r_0}{r+r_0}\Pi\Big(\tfrac{4rr_0}{(r+r_0)^2};\tfrac{\theta}{2}\Big|m\Big), \label{eq:Jtube}
	\end{align}
	where the definitions of $ L,L_0,m,A $ are the same as those of the cylinder.\\
	\indent Determining the correction terms in the same way as the case of cylinder, the electric (field line) potentials $ \phi^{\text{tube}}(r,z) $ and $ \psi^{\text{tube}}(r,z) $ made by a tube of radius $R$ and height $2Z$ with uniform surface charge density $ \sigma_0 $ is given by
	\begin{align}
		\phi^{\text{tube}}&=\sigma_0 R \sum_{\beta=\pm 1}2\beta I^{\text{tube}}(R,\pi,\beta Z-z;r), \label{eq:tub01} \\ 
		\psi^{\text{tube}}&=\sigma_0 R \sum_{\beta=\pm 1}2\beta J^{\text{tube}}(R,\pi,\beta Z-z;r) +\psi^{\text{tube}}_{\text{corr}},  \label{eq:tub02}
	\end{align}
	where
	\begin{align}
		\psi^{\text{tube}}_{\text{corr}} = 4\pi\sigma_0 RZ \operatorname{sgn}(z)\, H(R-r). \label{eq:tub03}
	\end{align}
	The plots are shown in Fig.~\ref{fig:tube}. Since the region with $ \rho=0 $ where $ \psi $ is definable is not simply connected, we can observe its topological character. That is, $ \psi $ is a multivalued function possessing the log-like branch structure, and the jump value between the neighboring branches
	\begin{align}
		\Delta \psi^{\text{tube}}= 8\pi R Z \sigma_0 = 2Q \label{eq:tub04}
	\end{align}
	can be thought of as a topological charge, where $ Q $ is a total electric charge of the tube. Thus, $ \psi^{\text{tube}}+n\Delta \psi^{\text{tube}} $ with $ n\in\mathbb{Z} $ becomes a single-valued function on the extended plane made by cut-and-glue of right half planes [Fig.~\ref{fig:tube} (b)]. 
	Accidentally, the topological and electric charges have the same dimension now. Indeed, if $ 4\pi \epsilon_0=1 $, the electric potential has the dimension $ [\phi]=\Big[\int\frac{\rho d\boldsymbol{r}'}{|\boldsymbol{r}-\boldsymbol{r}'|}\Big]= \mathrm{C}/\mathrm{m} $, and hence $ [\psi]=[r\phi]= \mathrm{C} $. \\
	\indent On the other hand, the topological discussion is obscured for cylinder, since the field line potential $ \psi $ expressed by Eq.~(\ref{eq:psidef}) is available only in a simply connected region [Fig.~\ref{fig:cyl} (b)]. Even if we numerically extrapolate the electric field lines inside the cylinder, their topology will be the same as those created by a charged ball, and therefore $\psi$ is still expected to be single-valued in the right half plane $(r\in\mathbb{R}_{>0},z\in\mathbb{R})$.  \\
	\indent The value of the potential on the surface $ r=R $ is of special interest. It is calculated via $ I_{\text{hyg}}(m,\sqrt{1-m};\pi) $, which corresponds to the boundary $|x|+|y|=1$ where the series (\ref{eq:AppellF2}) converges \cite{ErdelyiHTF1,KimuraHFTV1973}. It may be rewritten in several forms:\cite{Foot02}
	\begin{align}
		&I_{\text{hyg}}(m,\sqrt{1-m};\pi)=\pi\sqrt{1-m}F_2^{\text{Appell}}\left(\begin{smallmatrix}\frac{1}{2};\frac{1}{2},1 \\ 1,\frac{3}{2} \end{smallmatrix};m,1-m\right) \notag \\
		&=\int_m^1\frac{K(m)dm}{m\sqrt{1-m}} %= -\int^{m/(m-1)}_{-\infty}\frac{K(m)d\mu}{\mu} \notag \\
		=\left[ -\frac{\pi\mu}{8} F^4_3\left( \begin{smallmatrix}1,1,\frac{3}{2},\frac{3}{2} \\ 2,2,2 \end{smallmatrix};\mu \right)-\frac{\pi}{2}\ln \frac{-\mu}{16}  \right]_{\mu=m/(m-1)}, \label{eq:F43expression}
	\end{align}
	where $ F^p_q\left( \begin{smallmatrix}a_1,\dots,a_p \\ b_1,\dots,b_q \end{smallmatrix};x \right)=\sum_{j=0}^\infty \frac{(a_1)_j\dots(a_p)_j}{(b_1)_j\dots(b_q)_j}\frac{x^j}{j!} $ is the Barnes (generalized) hypergeometric function.

\begin{figure}[t]
	\begin{center}
	\includegraphics[width=.85\linewidth]{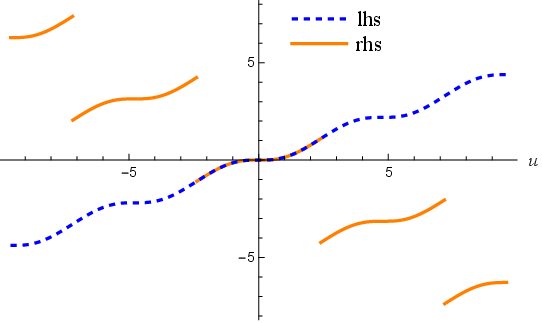}
	\caption{\label{fig:IntZSCbyAppell}Numerical check of Eq.~(\ref{eq:intZsc}) using AppellF2 in Mathematica 14.2. Here we set $ m=0.85 \ [K(m)\simeq 2.39] $ and the plot range is $ -4K(m)\le u \le 4K(m) $. The jump value at $ u=(2n+1)K(m), n\in\mathbb{Z} $, is given by $ \Delta=\frac{\pi}{2K(m)\sqrt{1-m}}\simeq 5.33. $}
	\end{center}
\end{figure}

	\indent \textit{Derivation of Eq.~(\ref{eq:IhygF2}) ---} Finally, we derive Eq.~(\ref{eq:IhygF2}). The naive expansion gives
	\begin{align}
		I_{\text{hyg}}=2As\sum_{l,j,k=0}^\infty \frac{(1)_j(\frac{1}{2})_{l+j}(\frac{1}{2})_k(\frac{1}{2})_{l+k}}{(\frac{3}{2})_j(\frac{3}{2})_{l+k}} \frac{(ms^2)^l (A^2)^j (s^2)^k}{l!j!k!}. \label{eq:rewriteI0}
	\end{align}
	According to \S 3.3 in Ref.~\citen{SrivastavaKarlsson}, this triple sum is written by the eleventh Lauricella-Saran function $ F_{11} (\text{or }F_M) $.
%	This triple sum does not reduce to the Lauricella function \cite{Exton}; we need a three-variable analog of the Kamp\'e de F\'eriet function for it. 
	Performing two of three summations, we obtain three different expressions:
	\begin{align}
		\!\!I_{\text{hyg}}\! &= 2As \sum_{l=0}^\infty \frac{(\frac{1}{2})_l}{2l+1}\frac{(ms^2)^l}{l!} F^2_1\left( \begin{smallmatrix} 1,\frac{1}{2}+l \\ \frac{3}{2} \end{smallmatrix};A^2 \right)F^2_1\left( \begin{smallmatrix} \frac{1}{2},\frac{1}{2}+l \\ \frac{3}{2}+l \end{smallmatrix};s^2 \right) \label{eq:rewriteI1} \\
		&=2As \sum_{j=0}^\infty  \frac{(\frac{1}{2})_j}{(\frac{3}{2})_j}(A^2)^j F_1^{\text{Appell}}\left( \begin{smallmatrix}\frac{1}{2}; \frac{1}{2}+j,\frac{1}{2} \\ \frac{3}{2} \end{smallmatrix}; ms^2,s^2 \right) \\
		&= 2As \sum_{k=0}^\infty  \frac{(\frac{1}{2})_k(\frac{1}{2})_k}{(\frac{3}{2})_k} \frac{(s^2)^k}{k!} F_2^{\text{Appell}}\left( \begin{smallmatrix} \frac{1}{2};\frac{1}{2}+k,1 \\ \frac{3}{2}+k,\frac{3}{2} \end{smallmatrix}; ms^2,A^2 \right),
	\end{align}
	where $ F^2_1\left( \begin{smallmatrix}\alpha,\beta \\ \gamma \end{smallmatrix};x \right) $ and $ F_1^{\text{Appell}}\left( \begin{smallmatrix}\alpha; \beta,\beta' \\ \gamma \end{smallmatrix};x,y \right) $ are Gauss's and Appell's hypergeometric function, respectively. None of them, however, can be used to obtain a summation-symbol-free expression for definite integral $ s=1 $. To this end, we apply the linear transformation 15.3.6 in Ref.~\citen{AbramowitzStegun} to the rightmost $ F^2_1 $ function in Eq.~(\ref{eq:rewriteI1}). Performing re-summation after this rewriting, we get Eq.~(\ref{eq:IhygF2}). Note that the sum in (\ref{eq:IhygF2}) also reduces to No. 10a of the table of \S 3.3 in Ref.~\citen{SrivastavaKarlsson}. \\
	\indent \textit{Integration of the product of Jacobi zeta and elliptic function ---} We further have a by-product formula as a bonus. Recalling another form of the elliptic integral of the third kind $ \Pi $ in terms of the Jacobi zeta and theta functions \cite{Akhiezer}, Eq.~(\ref{eq:IhygF2}) is re-interpreted as\cite{Foot02}
	\begin{align}
		&\int_0^u\! Z(u|m) \operatorname{sc}(u|m)du=-\operatorname{am}(u|m) \notag \\ 
		&\qquad +\frac{\pi\operatorname{sc}(u|m)}{2K(m)}F_2^{\text{Appell}}\left( \begin{smallmatrix}\frac{1}{2}; \frac{1}{2},1 \\ 1,\frac{3}{2} \end{smallmatrix}; m,(m-1)\operatorname{sc}^2(u|m) \right). \label{eq:intZsc}
	\end{align}
	Figure \ref{fig:IntZSCbyAppell} exhibits a numerical verification of Eq.~(\ref{eq:intZsc}), showing that this formula is valid in $ -K(m)< u < K(m) $, and the right hand side has jumps at $ u=(2n+1)K(m), \ n\in\mathbb{Z} $, whose value is found to be $ \Delta= \frac{\pi}{2K(m)\sqrt{1-m}} $ by the Euler transformation [\S 5.11 of Ref.~\citen{ErdelyiHTF1} and Eq. (6) in Ref.~\citen{Ananthanarayan2023}].\cite{Foot02} \\
	\indent Here, we comment on the significance of the formula (\ref{eq:intZsc}). Let $ \operatorname{JE} $ denote any of the twelve Jacobi elliptic functions. While the integral of $ Z \, \operatorname{JE}^2 $ is expressible by functions appearing in the classical theory of elliptic functions \cite{ByrdFriedman,AbramowitzStegun}, it is impossible for $Z\, \operatorname{JE}$. Therefore, providing formulas similar to Eq.~(\ref{eq:intZsc}) for all remaining Jacobi elliptic functions, it fills in the missing piece of the table of formulas.\cite{Foot01} 

	\indent \textit{Conclusion ---} We have presented the closed-form expressions of electric potentials and field lines for a uniformly-charged cylinder [Eqs.~(\ref{eq:cyl01})-(\ref{eq:cyl04})] and tube [Eqs.~(\ref{eq:tub01})-(\ref{eq:tub03})], expressed by elliptic integrals and Appell's hypergeometric functions. The field lines are plotted as equipotential lines of the field line potential (\ref{eq:psidef}), defined only in chargeless regions. The field line potential can have a multi-valued character if the chargeless region is not simply connected and the topological charge (\ref{eq:tub04}) can be introduced. The integration formula (\ref{eq:intZsc}) for the product of the Jacobi zeta and elliptic functions has also been presented, which was absent in classical tables of formulas.\\
	\indent The similar problem for conductors in electrostatic equilibrium would be worth investigating. Writing $ \psi $ in $ \rho\ne0 $ region seems challenging. Finding another nice application of special functions for other geometrical shapes remains open. 
%	\indent The same problem for low-dimensional objects including disk and ring is soon solved as a corollary. Writing $ \psi $ in $ \rho\ne0 $ region seems challenging. Finding another nice application of special functions for other geometrical shapes remains open. 

\begin{acknowledgment}
The author is grateful to the anonymous referee for drawing his attention to Ref.~\citen{SrivastavaKarlsson}, which improved the paper.
\end{acknowledgment}

%\appendix
%\section{}
%
%Use the \verb|\appendix| command if you need an appendix(es). The \verb|\section| command should follow even though there is no title for the appendix (see above in the source of this file).
%
%For authors of Invited Review Papers, the \verb|profile| command is prepared for the author(s)' profile.  A simple example is shown below.
%
%\begin{verbatim}
%\profile{Taro Butsuri}{was born in Tokyo, Japan in 1965. ...}
%\end{verbatim}

%\bibliographystyle{iopart-num}
%\bibliographystyle{unsrt}
%\bibliographystyle{ptephy_authornameperiod_nomonth}
\bibliographystyle{ptephy_authornameperiod_nomonth_noissuenumber_nolastpage}
\providecommand{\newblock}{}

%\bibliography{EPandEFLforCylandTube}

%\begin{thebibliography}{9}
%\bibitem{jpsj} The abbreviation for JPSJ must be ``J. Phys. Soc. Jpn." \note{in the reference list}.
%\bibitem{instructions} More abbreviations of journal titles are listed in ``Instructions for Preparation of Manuscript".
%\bibitem{etal} The use of ``et al.'' is not accepted in principle, therefore, all the authors must be listed.
%\bibitem{ibid} The term ``ibid.'' should not be used even if the same journal or book is cited with different page numbers.
%\bibitem{Errata} Errata should be listed under the same reference number. 
%\end{thebibliography}

%%%%%%%%%%%%%%%%%%%%%%%%%%%%%%%%%%%%%%%%%%%%%%%%%%%%%%%%%%%%%%%%%%%%%%%%%%%%%%%%%%%%%

%%%%%%%%%%%%%%%%%%%%%%%%%%%%%%%%%%%%%%%%%%%%%%%%%%%%%%%%%%%%%%%%%%%%%%%%%%%%%%%%%%%%%

\vspace{.5em}
\noindent (The Addendum and the Supplementary calculations are given in the remaining pages.)

\fancyfoot[C]{\footnotesize \thepage}

\clearpage
%\section{}
\makeatletter
\renewcommand{\@biblabel}[1]{#1)}
\makeatother

\thispagestyle{fancy}
\rhead{\textbf{\footnotesize ADDENDA}}
\lhead{\footnotesize J. Phys. Soc. Jpn.}

\begin{center}
\setlength{\baselineskip}{1.5em}
{\large\bf
 Addendum to ``Electric potentials and field lines for uniformly-charged tube and cylinder expressed by Appell's hypergeometric function and integration of $Z(u|m) \mathrm{sc}(u|m)$''}
\end{center}
\vspace{0.2em}
\begin{center} 
\setlength{\baselineskip}{1.1em}
 Daisuke A. Takahashi \\[.5em]
\textit{\small Research and Education Center for Natural Sciences, Keio University, Hiyoshi 4-1-1, Yokohama, Kanagawa 223-8521, Japan}
\end{center}
\vspace{2em}

% In the Addendum page appended to the main article, several relevant references are included and the decomposition of the solution by ``degree of transcendence'' is proposed.

\indent In Ref.~\citen{doi:10.7566/JPSJ.94.053001}, the exact solutions of the electrostatic potentials and the field line potentials written by elliptic integrals and Appell's hypergeometric functions have been provided.
The aim of this Addendum is (I) to provide relevant references which were missing in the previous paper and (II) to propose the decomposition of the potentials and the equations by their  \textit{``degrees of transcendence''}.\\

\indent (I) First, we mention the relevant literature. In Refs.~\citen{CIFTJA20122803,CIFTJA201845,Batle04052019}, electrostatic potentials and self energies for uniformly-charged  filled and hollow cylinders are calculated, and the self energy is expressed in terms of the generalized hypergeometric series ${}_4F_3$, the Appell, and the Kamp\'e de F\'eriet hypergeometric functions. Their calculations are based on expansion by orthogonal functions and many nontrivial integral formulae including Bessel functions have been derived.\\
\indent On analogous magnetic problems, the vector potentials and magnetic fields made by a finite-length solenoid is given by elliptic integrals \cite{osti_4121210} and generalization to inhomogeneous axisymmetric current distribution has been made with several examples \cite{Conway947050}. Their expression does not include multivariable hypergeometric functions, unlike the corresponding electric problems.\\
\indent Another interesting concept used in electrostatics (and equivalent classical gravity) named after Appell is the Appell ring \cite{Appell1887,RJGleiser_1989,Kofron:2023ndy}, which is defined as follows: If the charge and position parameters in the point-charge potential are formally changed from real to complex, the real part of the potential still satisfies the Laplace equation with vanishing property at spatial infinity. Therefore, it represents an electrostatic potential made by a certain charge distribution, which is referred to as the Appell ring.\\

\indent (II) Next, we report the decoupling property of the electric potential. The electric potential made by a uniformly-charged cylinder has been written by a sum of three terms, $ \phi^{\text{cyl}}=\phi^{\text{cyl}}_{\text{hyg}}+\phi^{\text{cyl}}_{\text{ell}}+\phi^{\text{cyl}}_{\text{\vphantom{hl}corr}} $ [see Eqs. (\ref{eq:cyl01}) and (\ref{eq:phicylcorr}) in Ref.~\citen{doi:10.7566/JPSJ.94.053001}]. In fact, we can further prove that they \textit{individually} satisfy the Laplace/Poisson equations except for the boundaries $r=R,\ z=\pm Z$, i.e., 
\begin{align}
	\left( \partial_r^2+\tfrac{1}{r}\partial_r+\partial_z^2 \right)\phi_{\text{corr}}^{\text{cyl}}&=-4\pi\rho_0 H(R-r)H(Z-|z|), \tag{A1} \\
	\left( \partial_r^2+\tfrac{1}{r}\partial_r+\partial_z^2 \right)\phi_{\text{T}}^{\text{cyl}}&=0, \tag{A2}
\end{align}
where $ \text{T}=\text{ell and hyg} $. 
Since  $ \phi_{\text{corr}}^{\text{cyl}},\ \phi_{\text{ell}}^{\text{cyl}}, $ and $ \phi_{\text{hyg}}^{\text{cyl}} $ are solely written by elementary functions, elliptic integrals, and Appell's hypergeometric function, this result implies that the equations are decoupled by the kinds of the special functions, or the \textit{``degrees of transcendence''}. It is also remarkable that the contributions from the source charge are all incorporated into the elementary part, $ \phi_{\text{corr}}^{\text{cyl}} $.\\
\indent If we could construct a theory such that the solutions of various partial differential equations and eigenvalue problems in physics are decomposed by such ``degrees of transcendence'', it could be a powerful tool. To refine this argument, we will need an appropriate mathematical definition for the degree of transcendence.

\bibliographystyle{ptephy_authornameperiod_nomonth_noissuenumber_nolastpage}
\providecommand{\newblock}{}

%\bibliography{EPandEFLforCylandTube}

%%%%%%%%%%%%%%%%%%%%%%%%%%%%%%%%%%%%%%%%%%%%%%%%%%%%%%%%%%%%%%%%%%%%%%%%%%%%%%%%%%%%%
% After pasting the content of bbl, 
% Add [An] like \bibitem[A1], \bibitem[A2], ... by hand! 
%(The ordinary method, i.e., renewcommand of biblabel does not work in jpsj3.cls... I could not find a solution.)

%%%%%%%%%%%%%%%%%%%%%%%%%%%%%%%%%%%%%%%%%%%%%%%%%%%%%%%%%%%%%%%%%%%%%%%%%%%%%%%%%%%%%

\vspace{1em}
\noindent (The Supplementary calculations follow next.)

\clearpage
%\section{}
\makeatletter
\renewcommand{\@biblabel}[1]{#1)}
\makeatother

\setcounter{secnumdepth}{0}

\makeatletter
\renewcommand{\theequation}{%
S\arabic{equation}}
%\thesection.\arabic{equation}}
\@addtoreset{equation}{section}
\makeatother

\fancyfoot[C]{\scriptsize \thepage}
\pagestyle{fancy}
\renewcommand{\headrulewidth}{0.0pt}
\rhead{}
\lhead{}

\setcounter{equation}{0}
\clearpage
\onecolumn

\begin{center}
{\large Supplementary calculations for J. Phys. Soc. Jpn. \textbf{94}, 053001 (2025) \\[.5ex] [arXiv:2503.11693, Jxiv:1133]} \\[1ex]
{\normalsize Daisuke A. Takahashi} \\[.5ex]
2025 Oct. 10
\vspace{1em}
\end{center}
%\twocolumngrid

\small
\setlength{\baselineskip}{1.5em}
*The supplementary calculations given here have been added for interested readers after publication. Hence, this is not an official supplemental material associated with a peer-reviewed paper.

\indent Here, we supply a detailed derivation of several expressions of the main article and explain the resemblance between the field line potential for tube and the electric potential for disk. We present the following contents:

\begin{enumerate}[(i)]\setlength{\itemsep}{-.25ex}
\item Derivation of Eq.~(\ref{eq:F43expression}), 
\item Derivation of the integration formula (\ref{eq:intZsc}),
\item Derivation of the jump value in Fig.~\ref{fig:IntZSCbyAppell},
\item Note on the identity of the elliptic integral of third kind given in the text after Eq.~(\ref{eq:Jcylell}),
\item Resemblance between the field line potential for \textit{tube} and the electric potential for \textit{disk}.
\end{enumerate}
% the derivation of the integral of $ Z(u|m) \operatorname{sc}(u|m) $.\\  %[arXiv:2503.11693, Jxiv:1133], including the derivation of the integral of $ Z \operatorname{sc} $.\\ 
\indent Unless otherwise noted, we follow the notations of elliptic integrals and functions in Ref.~\citen{AbramowitzStegun} and Mathematica.
For brevity, we write the integral (\ref{eq:Ihygdef}) as $I_{\text{hyg}}=I$: 
	\begin{align}
		I(m,A;\theta)\coloneqq \int_0^\theta\! d\theta\tanh^{-1}\frac{A}{\sqrt{\smash[b]{1-m\sin^2\frac{\theta}{2}}}}. \label{eq:Ihyg}
	\end{align}
The parameter derivatives are easily calculated:
	\begin{align}
		\frac{\partial I}{\partial A}&=2F\big(\tfrac{\theta}{2}|m\big)+\frac{2A^2}{1-A^2}\Pi\Big(\tfrac{2A^2}{1-A^2};\tfrac{\theta}{2}\Big|m\Big),  \label{eq:IA} \\
		\frac{\partial I}{\partial m}&=\frac{A}{m}\left[ \frac{2A^2}{1-A^2}\Pi\Big(\tfrac{2A^2}{1-A^2};\tfrac{\theta}{2}\Big|m\Big)-F\big(\tfrac{\theta}{2}|m\big) \right], \label{eq:Im}
	\end{align}
where $ F(\varphi|m) $ and $ \Pi(n;\varphi|m) $ are the incomplete elliptic integral of the first and the third kind.

\section{\small (i) Derivation of Eq.~(\ref{eq:F43expression})}
\indent Let us derive Eq.~(\ref{eq:F43expression}), which is used to calculate the value of electric potentials on the surface $r=R$.\\
\indent  We write the definite integral for $ \theta=\pi $ as $ I(m,A;\pi)=I(m,A) $. Then, the above parameter derivatives reduces to the complete elliptic integrals $ F\big(\tfrac{\pi}{2}|m\big)=K(m) $ and $ \Pi\Big(\tfrac{2A^2}{1-A^2};\tfrac{\pi}{2}\Big|m\Big)=\Pi\Big(\tfrac{2A^2}{1-A^2}\Big|m\Big) $. With $A=\sqrt{1-m}$, 
	\begin{align}
		\frac{\partial }{\partial m}I(m,\sqrt{1-m})=\frac{\partial I}{\partial m}\bigg|_{A=\sqrt{1-m}}+\frac{\partial \sqrt{1-m}}{\partial m}\frac{\partial I}{\partial A}\bigg|_{A=\sqrt{1-m}}=\frac{-K(m)}{m\sqrt{1-m}}.
	\end{align}
	Taking into account  $ I(1,0)=0 $, we re-integrate the above expression. With the modular transformation $ \mu=\frac{m}{m-1} $,
	\begin{align}
		I(m,\sqrt{1-m})=\int_m^1\frac{K(m)dm}{m\sqrt{1-m}} = -\int^{m/(m-1)}_{-\infty}\frac{K(\mu)d\mu}{\mu}, \label{eq:Ssur2}
	\end{align}
	where we have used $K\big(\frac{m}{m-1}\big)=\sqrt{1-m}K(m)$. (The list of such transformation formulae for elliptic integrals under the replacement of $m$ by some of the \textit{sextuple} $ \{m,\tfrac{m}{m-1},1-m,\tfrac{1}{m},1-\tfrac{1}{m},\tfrac{1}{1-m} \} $, which arises from the $\mathrm{PSL}(2,\mathbb{Z})$ transformation of the modular lambda function $m=\lambda(\tau)$ [e.g., Ref.~\citen{Akhiezer}, \S23], is found in Ref.~\citen{ErdelyiHTF2}, \S 13.7, TABLE 3.) Rewriting $K(\mu)$ as a hypergeometric series, the rightmost expression of (\ref{eq:Ssur2}) reduces to the generalized hypergeometric function, the last line of Eq.~(\ref{eq:F43expression}).
%	\begin{align}
%		I(m,\sqrt{1-m})=\left[ -\frac{\pi\mu}{8} F^4_3\left( \begin{smallmatrix}1,1,\frac{3}{2},\frac{3}{2} \\ 2,2,2 \end{smallmatrix};\mu \right)-\frac{\pi}{2}\ln \frac{-\mu}{16}  \right]_{\mu=m/(m-1)}.
%	\end{align}

\section{\small (ii) Derivation of the integration formula (\ref{eq:intZsc})}
	Introducing the new variable $ \sin\frac{\theta}{2}=\operatorname{sn}(u|m) $ (i.e., $ \theta=2\operatorname{am}(u|m) $), Eqs. (\ref{eq:Ihyg}) and (\ref{eq:IA}) become
	\begin{align}
		I&=2\int_0^u\!\!du \operatorname{dn}(u|m)\tanh^{-1}\frac{A}{\operatorname{dn}(u|m)}, \\
		\frac{\partial I}{\partial A}&= 2\int_0^u \!\! du \frac{\operatorname{dn}^2u}{-A^2+\operatorname{dn}^2u}. \label{eq:IA2}
	\end{align} 
	We parametrize $ A=-\mathrm{i}\sqrt{1-m}\operatorname{sc}(\mathrm{i}a|m)=\operatorname{dn}(\mathrm{i}a-K-\mathrm{i}K'|m) $, where $ K=K(m),\ K'=K(1-m) $. These parameters can be complex-valued, but if we want to keep physical context of the electrostatic problem with $m=\frac{4rr_0}{L_0^2}$ and $A=\frac{z}{L_0}$, the range of parameters should be $ -1 \le \frac{A}{\sqrt{1-m}} \le 1 $ and hence $ -K' \le a \le K' $. Henceforth we omit the second variable $m$ in elliptic functions. Rewriting the integrand in Eq.~(\ref{eq:IA2}) as $ 1-\frac{1}{m}\frac{\operatorname{dn}^2(\mathrm{i}a-K-\mathrm{i}K')}{\operatorname{sn}^2u-\operatorname{sn}^2(\mathrm{i}a-K-\mathrm{i}K')} $ and using the formula in Ref.~\citen{Akhiezer}, Chap.~5, \S 29 (or also Ref.~\citen{PhysRevE.93.062224}, Eq.~(B9)), we find
	\begin{align}
		\frac{\partial I}{\partial A}=2u+\frac{2\operatorname{sn}(\mathrm{i}a)\operatorname{cn}(\mathrm{i}a)}{\operatorname{dn}(\mathrm{i}a)}\left( \frac{1}{2}\ln\frac{\Theta_1(\mathrm{i}a-K-\mathrm{i}K'-u)}{\Theta_1(\mathrm{i}a-K-\mathrm{i}K'+u)}+uZ(\mathrm{i}a-K-\mathrm{i}K') \right),
	\end{align}
	where we use the scaled theta function (\ref{eq:scaledtheta}) introduced below.
	In particular, the definite integral $ u=K \ (\leftrightarrow \theta=\pi) $ is
	\begin{align}
		\left.\frac{\partial I}{\partial A}\right|_{u=K}=2K+\frac{2\operatorname{sn}(\mathrm{i}a)\operatorname{cn}(\mathrm{i}a)}{\operatorname{dn}(\mathrm{i}a)}\left( \frac{1}{2}\ln(-1)+KZ(\mathrm{i}a-K-\mathrm{i}K') \right). \label{eq:DIDA2}
	\end{align}
	We should choose $\ln (-1)=-\mathrm{i}\pi$ in Eq.~(\ref{eq:DIDA2}), because $ A=a\sqrt{1-m}+O(a^3) $ and hence $ \left.\frac{\partial^2I }{\partial A^2}\right|_{u=K}=0 $, implying that $ a=0 $ is a second-order zero of $ \left.\frac{\partial I}{\partial A}\right|_{u=K} $. Then, rewriting integrand using Eq.~(\ref{eq:zetaperiodicity04}), and noting that $ A=-\mathrm{i}\sqrt{1-m}\operatorname{sc}(\mathrm{i}a) $ implies $ dA=\sqrt{1-m}\operatorname{nc}(\mathrm{i}a)\operatorname{dc}(\mathrm{i}a)da $, the integral reduces to the form 
	\begin{align}
		I|_{u=K}=\int_0^A dA \left.\frac{\partial I}{\partial A}\right|_{u=K}=2KA+2K\sqrt{1-m}\int_0^a da \operatorname{sc}(\mathrm{i}a)\Big(Z(\mathrm{i}a)-\operatorname{sc}(\mathrm{i}a)\operatorname{dn}(\mathrm{i}a)\Big).
	\end{align}
	Since $ \operatorname{sc}^2\operatorname{dn}=\frac{(1-\operatorname{cn}^2)\operatorname{dn}}{\operatorname{cn}^2}=\operatorname{dc}\operatorname{nc}-\operatorname{dn}=\operatorname{sc}'-\operatorname{am}' $, and recalling that $I|_{u=K}$ is given by $[\text{Eq.~(\ref{eq:IhygF2})}]_{s=1}$, we find
	\begin{align}
		I|_{u=K}=-\mathrm{i}\pi\sqrt{1-m}\operatorname{sc}(\mathrm{i}a)F_2^{\text{Appell}}\left( \begin{smallmatrix}\frac{1}{2};\frac{1}{2},1 \\ 1,\frac{3}{2} \end{smallmatrix}; m,(m-1)\operatorname{sc}^2(\mathrm{i}a) \right)=-2\mathrm{i}K\sqrt{1-m}\left( \operatorname{am}(\mathrm{i}a)+\int_0^{\mathrm{i}a}dv \operatorname{sc}(v)Z(v) \right),
	\end{align}
	which reduces to Eq.~(\ref{eq:intZsc}) by replacement $ \mathrm{i}a \to u $.

{\scriptsize
\section{\scriptsize * Appendix to (ii) : notations of theta functions and (quasi-)periodicity of the Jacobi zeta function}
%
%footote begins%
%\footnote{
Here we use the scaled theta functions
\begin{align}
	\Theta_i(u|m)\coloneqq [\vartheta_i(\tfrac{\pi u}{2K(m)},q)]_{\text{Abramowitz-Stegun}},\quad i=1,2,3,4 \label{eq:scaledtheta}
\end{align}
with nome $q=\mathrm{e}^{-\pi\frac{K'}{K}}$, where the notation $ \vartheta_i(z,q) $ is based on 16.27 of Ref.~\citen{AbramowitzStegun}.
 $\vartheta_4(z,q)$ is sometimes also written as $\vartheta_0(z,q)$ (e.g., Refs.~\citen{Akhiezer,Todaelliptic}).  In Akhiezer's book\cite{Akhiezer}, \S 25, $ \Theta_i(u) $ is denoted by $ \theta_i(u) $ (\textit{not} vartheta). The correspondence between Eq.~(\ref{eq:scaledtheta}), Jacobi's original notations (Ref.~\citen{AbramowitzStegun}, 16.31 and Ref.~\citen{WhittakerWatson}, \S21.62), and Ref.~\citen{Akhiezer} is 
\begin{align}
	(\Theta_1,\Theta_2,\Theta_3,\Theta_4)_{\text{defined here}}=(H,H_1,\Theta_1,\Theta)_{\text{Jacobi}}=(\theta_1,\theta_2,\theta_3,\theta_0)_{\text{Akhiezer}}.
\end{align}
The Jacobi zeta function is written by (Ref.~\citen{AbramowitzStegun}, 16.34, 17.4.28)
\begin{align}
	Z(u|m)=\int_0^u du \left( \operatorname{dn}^2(u|m)-\frac{E(m)}{K(m)} \right)= \frac{\partial }{\partial u}\ln \Theta_4(u|m).
\end{align}
Using the quasiperiodicity
\begin{align}
	\Theta_4(v+2lK+2n\mathrm{i}K')=(-1)^{n}\mathrm{e}^{n^2\pi\frac{K'}{K}}\mathrm{e}^{-\frac{2n\mathrm{i}\pi v}{2K}}\Theta_4(v),\ l,n\in\mathbb{Z},
\end{align}
and the addition formulas of the Jacobi zeta function found in, e.g., Ref.~\citen{Akhiezer}, \S27 and Ref.~\citen{Todaelliptic}, p.201 (Note: in Toda's book\cite{Todaelliptic}, p. 201, Eq.~(\ref{eq:zetaadd2}) has a misprint; the numerator of the RHS should be modified as $ \operatorname{sn}^2u \to \operatorname{sn}u $), 
	\begin{align}
		Z(u\pm v)=Z(u)\pm Z(v)\mp m\operatorname{sn}u\operatorname{sn}v\operatorname{sn}(u\pm v), \label{eq:zetaadd1} \\
		Z(u+v)+Z(u-v)-2Z(u)=\frac{-2m\operatorname{sn}u\operatorname{cn}u\operatorname{dn}u\operatorname{sn}^2v}{1-m \operatorname{sn}^2u \operatorname{sn}^2v}, \label{eq:zetaadd2}
	\end{align}
 we can prove the quasiperiodicity of the Jacobi zeta function ($l,n\in\mathbb{Z}$):
	\begin{align}
		Z(v+2lK+2n\mathrm{i}K')&=Z(v)-\frac{\mathrm{i}\pi n}{K}, \\
		Z(v+(2l+1)K+2n\mathrm{i}K')&=Z(v)-\frac{m\operatorname{sn}v\operatorname{cn}v}{\operatorname{dn}v}-\frac{\mathrm{i}\pi n}{K}, \\
		Z(v+2lK+(2n+1)\mathrm{i}K')&=Z(v)+\frac{\operatorname{cn}v\operatorname{dn}v}{\operatorname{sn}v}-\frac{\mathrm{i}\pi(2n+1)}{2K}, \\
		Z(v+(2l+1)K+(2n+1)\mathrm{i}K')&=Z(v)-\frac{\operatorname{sn}v\operatorname{dn}v}{\operatorname{cn}v}-\frac{\mathrm{i}\pi (2n+1)}{2K}. \label{eq:zetaperiodicity04}
	\end{align}
	The analogous formulas are found in Ref.~\citen{AbramowitzStegun}, 16.34.1-16.34.4. (Note: the sign of the second term in the RHS of 16.34.3 should be $+m \frac{\operatorname{sn}u\operatorname{cn}u}{\operatorname{dn}u}$.)
%}
%footnote ends%

}

\section{\small (iii) Derivation of the jump value in Fig.~\ref{fig:IntZSCbyAppell}}
By the transformation given by Ref.~\citen{ErdelyiHTF1}, \S 5.11, Eq. (6), or Ref.~\citen{Ananthanarayan2023}, Eq.~(6),
\begin{align}
	F_2^{\text{Appell}}\left( \begin{smallmatrix}\alpha;\beta,\beta' \\ \gamma,\gamma'\end{smallmatrix}; x,y \right)=(1-x)^{-\alpha}F_2^{\text{Appell}}\left( \begin{smallmatrix}\alpha;\gamma-\beta,\beta' \\ \gamma,\gamma'\end{smallmatrix}; \tfrac{x}{x-1},\tfrac{y}{1-x} \right),
\end{align}
we find
\begin{align}
	\operatorname{sc}(u) F_2^{\text{Appell}}\left( \begin{smallmatrix}\frac{1}{2};\frac{1}{2},1 \\ 1,\frac{3}{2}\end{smallmatrix};m,(m-1)\operatorname{sc}^2(u) \right)=\operatorname{sc}(u) |\operatorname{cd}(u)|F_2^{\text{Appell}}\left( \begin{smallmatrix}\frac{1}{2};\frac{1}{2},\frac{1}{2}\\ 1,\frac{3}{2}\end{smallmatrix};m\operatorname{cd}^2u, (1-m)\operatorname{sd}^2 u \right). \label{eq:AppellF2ET}
\end{align}
Then, the limit $ \text{[Eq.~(\ref{eq:AppellF2ET})]} \overset{u\to K \pm 0}{\longrightarrow} \frac{\mp 1}{\sqrt{1-m}}F_2^{\text{Appell}}\left( \begin{smallmatrix}\frac{1}{2};\frac{1}{2},\frac{1}{2}\\ 1,\frac{3}{2}\end{smallmatrix};0,1 \right)=\frac{\mp1}{\sqrt{1-m}}\frac{\pi}{2} $ explains the jump value $\Delta=\frac{\pi}{2K(m)\sqrt{1-m}}$ in Fig.~\ref{fig:IntZSCbyAppell}.

\fancyfoot[C]{\scriptsize \raisebox{8pt}{\thepage}}
\pagestyle{fancy}

\section{\small (iv) Note on the identity of the elliptic integral of third kind given in the text after Eq.~(\ref{eq:Jcylell})}
Here, we provide a comment on the formula in the text between Eqs. (\ref{eq:Jcylell}) and (\ref{eq:cyl01})
\begin{align}
	\sum_{\alpha=\pm}\left[ 1\!-\!\tfrac{n_\alpha}{2}\left( 1\!+\!\tfrac{r}{r_0} \right) \right]\Pi(n_\alpha|m)=K(m)+\frac{r-r_0}{r+r_0}\Pi \Big(\tfrac{4rr_0}{(r+r_0)^2}\Big|m\Big)+\frac{\pi L_0}{|z|}H(r_0-r), \label{eq:Piformula}
\end{align}
where $ L_0=\sqrt{(r+r_0)^2+z^2},\  m=\frac{4rr_0}{L_0^2},\ A=\frac{z}{L_0},\ n_\pm=\frac{2r_0}{r_0\pm\sqrt{\smash[b]{r_0^2+z^2}}} $, and $ H(x) $ is the Heaviside step function. Equation~(\ref{eq:Piformula}) can be found by comparing two expressions for the electrostatic potential made by a uniformly-charged disk derived based on our calculation and that by Lass and Blitzer \cite{Lass1983}. Below, we discuss it in a little more detail.\\
\indent The double indefinite integral used for the electric potential made by a uniformly-charged disk is
\begin{align}
	&I^{\text{disk}}(r,\theta,z;r_0)=\frac{\partial I^{\text{cyl}}(r,\theta,z;r_0)}{\partial z}=\iint \frac{rdrd\theta}{L}\notag \\
	&=-r_0\sin\theta \tanh^{-1}\frac{r+r_0\cos\theta}{L}+L_0E(\tfrac{\theta}{2}|m)+\frac{r^2-r_0^2-z^2}{L_0}F(\tfrac{\theta}{2}|m)+\frac{z^2}{L_0}\sum_{\alpha=\pm}\left[ 1-\tfrac{n_\alpha}{2}\Big(1+\tfrac{r}{r_0}\Big) \right]\Pi(n_\alpha;\tfrac{\theta}{2}|m)
\end{align}
with $L=\sqrt{r^2+r_0^2+2rr_0\cos\theta+z^2}$. Then, the electric potential made by the disk of radius $R$ with unit charge density is given by $ \phi^{\text{disk}}(r,z)=2(I^{\text{disk}}(R,\pi,z;r)-I^{\text{disk}}(0,\pi,z;r)) $, yielding
\begin{align}
	\phi^{\text{disk}}(r,z)=\frac{2}{L_0}\left( L_0^2 E(m)+(R^2-r^2-z^2)K(m)+z^2\sum_{\alpha=\pm}\left[ 1-\tfrac{n_\alpha}{2}\Big(1+\tfrac{r}{r_0}\Big) \right]\Pi(n_\alpha|m)  \right)-2\pi |z|. \label{eq:phidisktak}
\end{align}
with $L_0=\sqrt{(R+r)^2+z^2},\ m= \frac{4rR}{L_0^2},\ A=\frac{z}{L_0}, n_\pm=\frac{2r}{r\pm\sqrt{\smash[b]{r^2+z^2}}}$.\\
\indent On the other hand, Lass and Blitzer\cite{Lass1983} have calculated the same potential using the different coordinate, whose origin is located at a point obtained by the projection of the observation point onto the disk. The resultant expression is
\begin{align}
	\phi^{\text{disk}}(r,z)= \frac{2}{L_0}\left( L_0^2 E(m)+(R^2-r^2)K(m)-\frac{r-R}{r+R}z^2\Pi\Big(\tfrac{4rR}{(r+R)^2}|m\Big) \right)-2\pi|z| H(R-r). \label{eq:phidiskLB}
\end{align}
Equating the two expressions (\ref{eq:phidisktak}) and (\ref{eq:phidiskLB}), and changing the letters $(R, r) \to (r, r_0) $, we obtain Eq.~(\ref{eq:Piformula}).\\
\indent It seems to be difficult to directly prove Eq.~(\ref{eq:Piformula}) without relying on physical context and only using the known formulas for the elliptic integral of the third kind given in Ref.~\citen{ByrdFriedman}. The incomplete analog of Eq.~(\ref{eq:Piformula}) is also unknown.

\section{\small (v) Resemblance between the field line potential for tube and the electric potential for disk}
Can we derive the Lass-Blitzer expression (\ref{eq:phidiskLB}), which is obviously simpler than (\ref{eq:phidisktak}), without finding the coordinate used in Ref.~\citen{Lass1983}? 

To this end, it should be noted that Eq.~(\ref{eq:phidiskLB}) and the indefinite integral $J^{\text{tube}}$ [Eq. (\ref{eq:Jtube})], which is used for calculation of the field line potential made by uniformly-charged tube $\psi^{\text{tube}}$ [Eqs. (\ref{eq:tub02}) and (\ref{eq:tub03})], share the similar mathematical expressions. The reason of this resemblance can be explained by Eq. (\ref{eq:axialCR}). That is, 
\begin{align}
	\psi^{\text{tube}} \sim \int\!\! dr \, r \frac{\partial \phi^{\text{tube}}}{\partial z} = \int\!\! dr \,r \phi^{\text{ring}} = \phi^{\text{disk}}, \label{eq:psitubephidisk}
\end{align}
where ``$\sim$'' means ``equal up to the correction term in Eq. (\ref{eq:psidef})'', and  $ \phi^{\text{ring}} $ represents the electric potential made by a uniformly-charged one-dimensional ring with ignorable thickness. The correspondence (\ref{eq:psitubephidisk}) offers another derivation of the expression (\ref{eq:phidiskLB}) within a calculation based on the familiar cylindrical coordinate.

\bibliographystyle{ptephy_authornameperiod_nomonth_noissuenumber_nolastpage}
\providecommand{\newblock}{}

%\bibliography{EPandEFLforCylandTube}

\begin{thebibliography}{10}
\bibitem{Foot01}
A similar example of the classically inexpressible integrals of elliptic
  functions written by $F_1^{\text{Appell}}$ is the integral of
  $\operatorname{JE}^\alpha$ with noninteger $ \alpha $. It has already been
  implemented by Mathematica.
\end{thebibliography}

%%%%%%%%%%%%%%%%%%%%%%%%%%%%%%%%%%%%%%%%%%%%%%%%%%%%%%%%%%%%%%%%%%%%%%%%%%%%%%%%%%%%%
% After pasting the content of bbl, 
% Add [An] like \bibitem[A1], \bibitem[A2], ... by hand! 
%(The ordinary method, i.e., renewcommand of biblabel does not work in jpsj3.cls... I could not find a solution.)

\renewcommand{\refname}

%%%%%%%%%%%%%%%%%%%%%%%%%%%%%%%%%%%%%%%%%%%%%%%%%%%%%%%%%%%%%%%%%%%%%%%%%%%%%%%%%%%%%%%

\end{document}